\documentclass[lettersize,journal]{IEEEtran}
\usepackage{lipsum,amsmath,multicol}
\usepackage[left=3cm,right=2cm,top=2cm,bottom=2cm]{geometry}
\usepackage{amsfonts}
\usepackage{algorithmic}
\usepackage{algorithm}
\usepackage{array}
\usepackage[caption=false,font=normalsize,labelfont=sf,textfont=sf]{subfig}
\usepackage{textcomp}
\usepackage{stfloats}
\usepackage{url}
\usepackage{verbatim}
\usepackage{graphicx}
\usepackage{cite}
\usepackage{mathtools}
\usepackage{siunitx}
\usepackage{booktabs}
\usepackage{multirow}
\usepackage{makecell}
\usepackage{anyfontsize}
\usepackage{tikz}
\usepackage{tabularx}
\usepackage{tikz}

\graphicspath{ {./images/} }

\begin{document}
\title{Enhanced Generalized OFDM with Index Modulation}

\author{
\IEEEauthorblockN{A.Atef Ibrahim, Amr A.Nagy, Ashraf Mahran, Amr Abdelaziz }\\ \IEEEauthorblockA{Department of communication, Military Technical College, Cairo, Egypt.}\\

\IEEEauthorblockA{ahmedatef1982@hotmail.com,amr.a.nagy@mtc.edu.eg}}

\markboth{}%
{Shell \MakeLowercase{\textit{et al.}}: A Sample Article Using IEEEtran.cls for IEEE Journals}

\maketitle

\begin{abstract}
In recent years, many attempts have been made to enhance Orthogonal Frequency  Multiplexing with Index Modulation (OFDM-IM) in terms of spectral efficiency and error performance. Two challenges typically erupt when using OFDM-IM. First, the degradation in spectral efficiency due to the subcarrier's deactivation, especially when using higher order modulation (M-ary) where every inactive subcarrier will cost  $Log_2(M)$ bits loss. Second, using a fixed number of active subcarriers within a sub-block forces the error to be localized within the sub-block. Yet, it loses the advantage of exploiting all possible pattern combinations degrading the overall spectral efficiency. In this paper, we introduce a solution to tackle those problems. The Enhanced Generalized Index Modulation (EGIM) is a simple systematic way to generate and detect the OFDM-IM frame. Unlike the classical OFDM-IM generation by splitting the frame into sub-frames which increases the complexity of the OFDM-IM transmitter and reflects on the receiver Maximum likelihood detector, EGIM Makes full use of all possible combinations of active subcarriers within the frame by using variable active subcarriers (k) depending on the incoming data. The EGIM is still susceptible to error propagation if the OFF symbol is wrongly mapped to one of the ON symbols or vice versa.
For that reason, we offer an OFDM-IM autoencoder to overcome this problem. The encoder generates the (ON/OFF) symbols systematically to achieve the advantage of sending all possible frame indices patterns depending on the input bit stream offering an average of 3dB gain in terms of power efficiency. The proposed encoder performance was compared to the standard encoder with the same effective coding rate using soft and hard decision Viterbi decoding utilizing the power gain achieved.
\end{abstract}


\section{Introduction}
\IEEEPARstart{T}{he} idea of index modulation is first introduced in \cite{bacsar2013orthogonal} as a technique to enhance the spectral efficiency of multi-carrier communication systems over frequency-selective and rapidly time-varying fading channels, transmitting additional bits through activating and deactivating data subcarriers within data sub-block which creates a new domain of transmitting data (index domain). Dividing the OFDM frame into multiple sub-blocks  using low order modulation (BPSK) symbols boosts spectral efficiency to exceed that of the classical OFDM. Yet, there are some points to be investigated:   
\begin{enumerate}
 \item Performance evaluation of general M-ary signal constellations. 
\item Choosing the optimal value of active sub-carrier $k$ and total number of bits for each sub-block n.
\end{enumerate}
Both factors negatively affect the spectral efficiency of IM communication  systems as follows:
\begin{enumerate}
    \item The deactivation of the subcarrier using high order modulation will cause a loss of $\log_2(M)$ bits per subcarrier.
    \item The best selection of $(n,k)$ to obtain the maximum spectral efficiency of the communication system where $n$ is the number of bits within the subframe.%of combinations of patterns of indices for each data frame can be achieved by dealing with the total length of the frame N by setting $k=\frac{N}{2}$. 
\end{enumerate}
\par The work in \cite{bacsar2013orthogonal} tries to solve the problems that emerge in the early trials of using the index as a new domain for sending data as follows :
\begin{enumerate}
    \item In \cite{abu2009SIM} subcarrier indices used for data transmission with a high probability of error propagation if the receiver misses the correct indices positions.
    \item The error propagation problem is solved in \cite{tsonev2011ESIM} by using fixed $k=\frac{N}{2}$, this prevents the error from propagation outside the frame boundaries but leads to huge computational complexity at the receiver to detect which index pattern was sent.
\end{enumerate}
 %The highest possible spectral efficiency can be achieved by using higher order modulation $M$, as well as maximizing the number of possible indices pattern but,  the large number of combinations introduces a huge computational complexity for the receiver detector to determine the sent indices pattern\cite{tsonev2011enhanced}.
 \par Many problems are still challenging in OFDM-IM communication systems can be summarized as follows:
\\ \textbf{The complexity of the receiver detector}: This results from the large number of index patterns the receiver is required to distinguish( e.g.: in \cite{tsonev2011ESIM} there is $C^N_{\frac{N}{2}}$), the problem was solved by dividing the frame into many sub-frames so that the receiver has low allowable patterns to detect. However, this will limit spectral efficiency results of the system comes from the index pattern combinations term  $\lfloor\log_2{C^n_k } \rfloor $  due to the reduction in $C^n_k  $ function from N to n. \\\textbf{The optimal activation ratio for maximum spectral efficiency}: hard efforts have been made in \cite{IM-entropy} in the derivation of the closed form of maximum entropy for an OFDM-IM communication system. \\ \textbf{Using a variable number of active subcarriers over the whole frame length with high-order modulation}:  on one hand, all possible pattern combinations will be available for the index domain. on the other hand, due to the lack of receiver knowledge of the number of active indices, the error in determining the number and position of active subcarriers within the frame causes the errors to propagate across the OFDM-IM frames\cite{abu2009SIM}. 
\\ Finally, there is a critical need to enhance the generalization of the OFDM-IM schemes to accommodate all combination patterns while maintaining the low complexity of  ML detection at the receiver. Simultaneously, a comprehensive solution for the error propagation problem is crucial to allow the practical application of the OFDM-IM system.
\\ Different forms of IM  suited for MIMO and Multicarrier communication systems were mentioned in \cite{basar2016IM5G} and  \cite{bacsar2017IM5GB}. Additionally, an  outstanding work on the generalization of OFDM-IM was proposed in \cite{fan2015GIM} where two main techniques were proposed:
\begin{enumerate}

\item{GIM-1} used for BPSK where, the number of the active index is taken from a set of allowable number  K =$\{k_1,k_2,....,k_r\}$. where r is the size of the index set size, with extreme case K =$ \{0,1,....,n\}$ which enhances the spectral efficiency over the classical OFDM-IM.
\item{GIM-2} higher order modulation (QPSK) is implemented over OFDM-IM without loss of spectral efficiency by considering the in-phase and quadrature components as two independent BPSK streams. The index modulation is applied to the in-phase and quadrature data independently.
\end{enumerate}
 
Both techniques (GIM-1 and GIM-2) exceed the classical OFDM-IM in terms of spectral efficiency. Additionally, wherever the existence of equality in spectral efficiency the simulations show that the above-mentioned techniques outperform the classical OFDM-IM in bit error rate overall the entire signal-to-noise ratio (SNR). This comes over the expenses of the computational complexity in the log-likelihood ratio (LLR) or Maximum likelihood (ML) detector, the complex multiplication in classical OFDM-IM is $~O(M)$ per subcarrier. However, the complexity in GIM1 is $~O(rMn)$ and GIM2 is $~O(2M)$ per subcarrier. The authors didn't deal with the effect of error propagation of GIM-3 (a combination of GIM-1 and GIM-2) for errors in indices position caused by channel conditions. 
 \par Based on the previous concept the technique of OFDM with in-phase/quadrature index modulation (OFDM-I/Q-IM) was presented in \cite{zheng2015IQIM} and \cite{wen2017index}. It can be considered one of the best candidates for high-speed communications \cite{zhang2020polar}, where the activation of in-phase and quadrature components occurs independently resulting in higher spectral efficiency (index bits activate  Both I and Q components).
\\The work in \cite{wen2017enhanced} presented two additional schemes using higher-order modulation:
 \begin{enumerate}
 \item OFDM-HIQ-IM where the in-phase and quadrature components are jointly activated with index selection bits treated independently as two different streams sending different index patterns on the I and Q components.
\item LP-OFDM-IQ-IM  spreads information symbols across two adjacent active
subcarriers through linear constellation precoding $2\times2$ matrix to achieve
additional diversity gain.
\end{enumerate}
  Efforts in  \cite{yang2015spectrum} were made to enhance the spectral efficiency by using adaptive mapping for each picked-up value selected from a set of allowable active indices (e.g. for the low number of active indices choose higher order modulation mapping and vice versa).\\
In the DM-OFDM-IM technique in \cite{DM-OFDM-IM} and \cite{NEW-DM} the process of activation and deactivation of the subcarrier takes place by switching between two mapping schemes, the problem in this approach is that the minimum Euclidean distance between symbols is quietly reduced to a small value thus, more power has to invested to achieve reasonable error performance. Where the chance of error propagation still exists, especially in low SNR.
\par The most recent generalized scheme introduced in \cite{sengupta2024generalized} where each data sub-block has a different number of active subcarriers (k), associated with Modulation Mode (MM),
the active subcarrier number $k_r$  will be picked up from the $K$ set with size R. Where,\\$K =\{k_1,k_2,..,k_r,..,k_R\}
\forall\,1\leq r\leq R,1\leq k_r \leq n.$
\\For achieving the full degree of freedom of the OFDM-IM frame pattern combinations, it is required to involve all possible combinations of a pattern generated by the $ C^N_k  $ function, this is valid when the value of $ k=N/2 $,  this can be fulfilled by consider the k as a random variable whose Binomial distribution with mean $N/2 $. Due to the lack of knowledge of the number of active subcarriers, any error in determining the position of active subcarriers uploaded by higher-order modulation symbols will cause errors to propagate within the boundaries of the frame affecting all successive frames. 
 \par This paper generalizes the use of OFDM-IM over the whole OFDM frame without splitting. This takes advantage of the maximum indices pattern combinations and reduces the transmitter complexity where there are no look-up tables for index selection within the subframe. The presented EGIM uses a variable number of active subcarriers in each frame to involve all possible pattern combinations within the whole frame length. More importantly, EGIM can be considered as a pivotal solution to the error propagation problem that emerges when using variable number active indices, $k$, in OFDM-IM through a customized convolutional encoder whose output is mapped into (ON/OFF) symbols covering all possible pattern combinations while, in addition, taking advantage of its error-correcting capabilities.    

The contributions of this work can be summarized as follows:
\begin{enumerate}
\item A solution for using high-order modulation with a variable number of active indices over the whole  OFDM-IM  frame without a dramatic loss in spectral efficiency through the use of a variable number of active index k with average value $ \frac{N}{2} $, instead of using of fixed (n,k) pair.
\item The Design and performance evaluation of EGIM technique where two Case studies were introduced using a systematic way to generate and detect OFDM-IM without splitting and reassembling the frame, as it is implemented without extra complexity computation were added to both techniques (the same as the classical OFDM).
\item The design Of OFDM-IM autoencoder generates active and inactive symbols simultaneously utilizing convolutional encoder error capabilities to eliminate error propagation within OFDM-IM frames.  
\end{enumerate}

 \par  The two case studies of EGIM  are:\\
 a)\textbf{The EGIM-4QAM} where the input bit stream is converted into active and inactive symbols uploaded over OFDM frame depending on the input data. The transmitted symbols for active subcarriers are drawn from the normal 4QAM constellation and the inactive subcarrier is sent with zero power (i.e.: 0+j0).\\
b)\textbf{The EGIM-8PSK}  is a modified version of DM-OFDM introduced in \cite{DM-OFDM-IM}, where the two 4QAM schemes $(M_A, M_B)$ is replaced with 8PSK scheme for active subcarriers, also  OFF symbol inserted for inactive subcarriers.  
%\par The systematic method of generation and detection of OFDM-IM reduces the high complexity of classical OFDM-IM Maximum Likelihood (ML) detector required to detect  $\sum_{k=0}^NC^N_k$ possible pattern combinations. 

The rest of the paper is organized as follows, in section \ref{sec:OFDM-IM} a brief background of classical OFDM-IM is presented where the main idea of OFDM-IM is explained. In section \ref{sec:EGIM}, the proposed EGIM technique was depicted with a detailed illustration for its two case studies with the derivation of the spectral efficiency and symbol error performance formulas.   The OFDM-IM autoencoder is depicted in section \ref{sec:auto_enc}. Simulation results are presented and analyzed in section  \ref{sec:sim_res}. Finally, the conclusion of the paper is given in section \ref{sec: conclusion}.\\  
%In this paper, we will use boldface uppercase letters for matrices, uppercase letters for frequency-domain representation, lowercase letters for time-domain representation. While, $(.)^{*}$ denotes the conjugate of a complex number, $(.)^{H}$ denotes a conjugate transpose, $\mathbf{I}_N$ denotes an identity matrix of size $N$,%/ $\mathbf{tr}(.)$ denotes a matrix trace operator,  $\det(.)$ denotes a matrix determinant operator and diag(.) denotes the diagonal matrix. 
\section{Background on OFDM-IM}
\label{sec:OFDM-IM}
\begin{figure}[!t]
\centering
\includegraphics[width=3.2in]{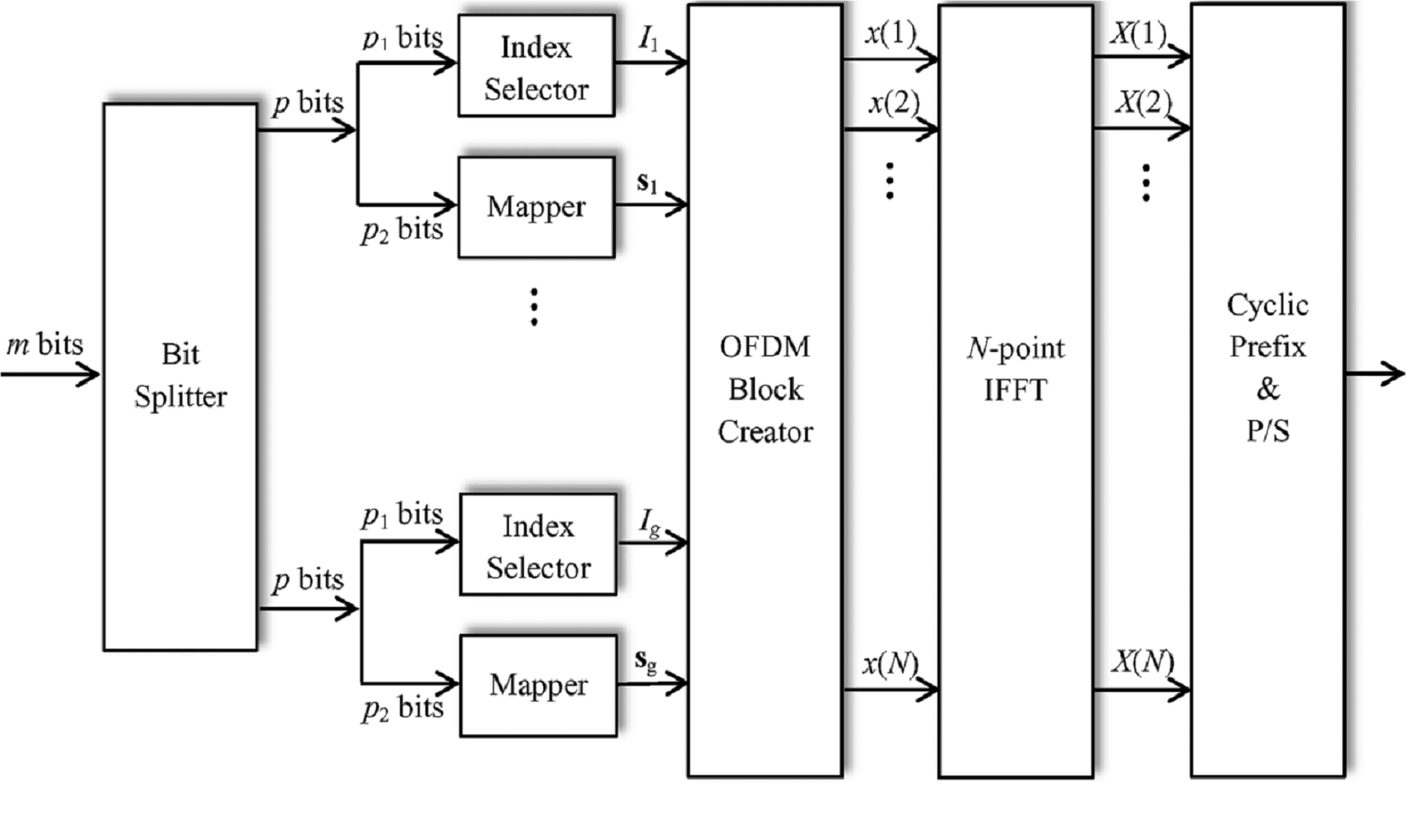}
\caption{Classical OFDM-IM transmitter introduced in \cite{bacsar2013orthogonal}.}
\label{fig:OFDM-IM}
\end{figure}
%\begin{list}{}{}
%\item{\url{http://www.latex-community.org/}} 
%\item{\url{https://tex.stackexchange.com/} }
%\end{list}

% 
  Fig.\ref{fig:OFDM-IM} shows the system model of classical OFDM-IM introduced in \cite{bacsar2013orthogonal}, where the input bit stream B is split into many groups G, i.e.: B=pG. Where p is the number of bits per group, similarly the OFDM frame length is divided into smaller groups N=nG. there is k number of active subcarrier within the group such that k$\le$n. The p bits are split into two sections i.e.: $p=p_1+p_2$, the first one $p_1$ is used for the index selection pattern I. However, the second $p_2$ is mapped to a symbol S uploaded on the active index. 

The maximum achievable bits gained by selecting k active index out of the number of bits in the group n are: \begin{equation}
B_1=p_1G= \lfloor\log_2{C^n_k } \rfloor G.
\end{equation}

While the maximum achievable bits gained by mapping bits over higher-order modulation
 \begin{equation}
B_2=p_2G= k(\log_2{M }) G.
\end{equation}
Resulting total achievable bits per OFDM-IM frame:  \begin{equation}\label{eq:SE_OFDMIM}
B=B_1+B_2=\lfloor\log_2{C^n_k } \rfloor G+ k\log_2{M } G.
\end{equation}

Recombine both types of bits to form OFDM-IM frame:
 \begin{equation}
X_F= [X(1),X(2),....,X(N)]^T.
\end{equation}
where $ X(\alpha)\in{0,S} $, is the subcarrier states either inactive or active the last one carries constellation symbol $\alpha$ can take values from {0,..,N}.
The frame $\mathbf{X}$ is processed as classical OFDM, taking the inverse fast Fourier transform (IFFT)as follows:
 \begin{equation}
X_T= \dfrac{N}{\sqrt{k}}IFFT(X_F)
\end{equation}
$X_T$ is the time domain representation of OFDM frame, $\dfrac{N}{\sqrt{k}}$ is the normalization factor to ensure that $\mathit{E}\{{X_T}^HX_T\}=N $, the receiver operate reversely where normalization factor $\dfrac{\sqrt{k}}{N} $is used in FFT process. 
\par One of the key challenges in the detection of OFDM-IM is that the receiver has to detect the active index positions first and then demodulate the symbols carried by those active subcarriers, error emerges when the receiver fails to detect the active indices accurately, it results loss in either pattern and constellation bits. This error can be bounded within OFDM-IM when using fixed values of $(n,k)$, otherwise (using variable active index each frame or subframe) the error in one frame will propagate across the following successive frames\cite{DM-OFDM-IM} cause error floor over the entire SNR range.

\section{Enhanced Generalized Index Modulation} \label{sec:EGIM}
The EGIM implementation of the IM concept by using a random input stream to determine the number and the index position of active subcarriers within the frame through the Bit stuffing technique which can be described simply as follows: 

 For a random binary input stream, the resulting output will be as follows:
\begin{list}{}{}
\item For "0" input bit the output will be $ 
\log_2{M }$ zeros.  
\item For "1" input bit the output will be the input bit followed by  $\log_2{M} $ bits. 
\end{list} 

\begin{figure}[!t]
\centering
\includegraphics[width=3.2in, height=2.2in]{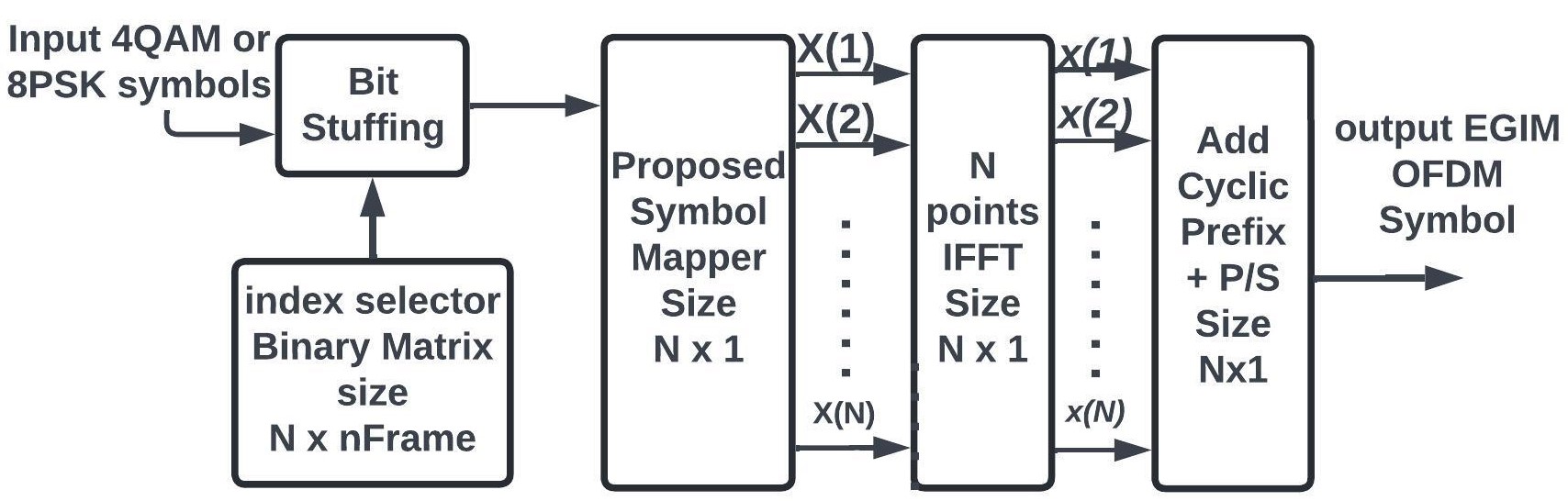}
\caption{Equivalent Transmitter of  Enhanced Generalized  OFDM-IM.}
\label{fig:Bit_stuffing_TX}
\end{figure}

The benefit of this technique is the reduction of error propagation probability, we will deduce the realization of such event later in this paper .\\
\textbf{Example:} the output for 1101010111100 input over QPSK will be:\\
Output 110 101 000 111 100.  
\par The above method is equivalent to using N bits stream for index selection where the number of the input binary "1" corresponds to the number and indices of the active subcarrier which carries symbols drawn from the 4QAM constellation as shown in Fig.\ref{fig:Bit_stuffing_TX} . In other words, the most significant bit in each output symbol corresponds to the index bit. 
The inactive subcarriers carry the off symbol, the original zero bit stuffed with $\log_2M$zeros (i.e.: "00")  to preserve the number of bits within a symbol. 
\subsection{Spectral Efficiency Analysis}\label{sec:SE_BS}
The number of bits  per frame that can be sent using QAM/PSK constellation is calculated  as follows :
\begin{equation}\label{eq:BS_SE}
B_1=k \log_2(M)
 \end{equation}
 For each frame, the MSB of each symbol represents the  N bits for subcarrier activation (index bit), the number of all possible combinations denoted by:
 \begin{equation}
     B_2= \lfloor \log_2{C^N_k}  \rfloor
 \end{equation}
  where k is a  random variable with Binomial  distribution with mean $\frac{N}{2}$ represents  the number of active  subcarriers  .\\The average spectral efficiency  per frame will be :
 \begin{equation}
     B_{EGIM}= E\{ \lfloor \log_2{C^N_k}\rfloor + k \log_2(M)\}
\end{equation}
\begin{equation}
 B_{EGIM} =\sum_{k=0}^{k=N}(\log_2{C^N_k}+(\frac{N}{2})\log_2{M})  
\end{equation}
Invoking the log sum inequality, we can write :
\begin{equation}
 B_{EGIM}  \geq \lfloor \log_2\sum_{k=0}^{k=N}{C^N_k}\rfloor +(\frac{N}{2})\log_2{M} 
\end{equation}

 The first term is the expansion for $2^N$ binomial and $\frac{N}{2}$ is the expected value of the active subcarrier random variable, thus the average spectral efficiency per subcarrier will be:
 \begin{equation}\label{eq:SE_BS}
  B_{EGIM}  \geq \frac{1}{N}\lfloor \log_22^N\rfloor +(\frac{1}{2})\log_2{M}   
\end{equation}
Therefore the final average spectral efficiency formula per subcarrier for bit stuffing is  :

 \begin{equation}\label{eq:SE-EGIM}
B_{EGIM}^{th}\geq  1+\frac{1}{2}\log_2{M}
 \end{equation}
\subsection{The First Case Study 4QAM }
\label{sec: EGIM-1}

\begin{figure}[!t]
\centering
\includegraphics[width=3.0in,height=6cm]{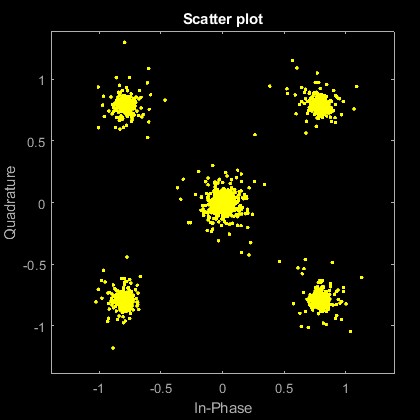}
\caption{EGIM-4QAM constellation over AWGN at SNR = 10 dB.}
\label{fig:Bitstuff_Const}
\end{figure}
  The output of the EGIM transmitter in the first case study is limited to  five codewords (symbols) are shown in Table \ref{tab:Bitstuff Mapper}, they are similar to the modified power-saving OFDM policy introduced in \cite{abu2009SIM} described in Fig.\ref{fig:Bitstuff_Const} 
 ,  a new fifth point is introduced to represent all zero symbols (inactive subcarrier) in addition to the $4QAM$ constellation. 
 \\Although the insertion of the $ 5^{th} $ constellation point (origin point) makes the proposed symbol mapper carry (3 bits) instead of (2 bits) in classical 4QAM. The drawback of this technique is that the minimum Euclidean distance between constellation points is reduced from $ \sqrt{2} $ unit to 1 unit ( assuming normalized power).
 \begin{table}[!t]
\caption{The proposed  Constellation for the first case study}
\label{tab:Bitstuff Mapper}
\centering
\begin{tabular}{|c||c||c|}
\hline
symbol&code word & constellation\\
\hline
$S_o$&000& $0+0j$\\
\hline
$S_1$&100& $\dfrac{1}{\sqrt{2}}(1+j)$\\
\hline
$S_2$&110& $\dfrac{1}{\sqrt{2}}(-1+j)$\\
\hline
$S_3$&111& $\dfrac{1}{\sqrt{2}}(-1-j)$\\
\hline
$S_4$&101& $\dfrac{1}{\sqrt{2}}(1-j)$\\
\hline
\end{tabular}
\end{table}
The result transmitted vector will be:
\begin{equation}
  \begin {split}
X&=\left[x_1,x_2,..,x_N\right] where\\ &\,x_i\in\{S_o,...,S_4\}\,and\, i=\left[1,..,N\right]
\end{split}  
\end{equation}
 For our proposed constellation shown in Fig.\ref{fig:Bitstuff_Const} the active carrier to be selected from constellation $M=4$ and inactive subcarrier carry zero bit, so the SE per subcarrier from (\ref{eq:SE-EGIM}) will be $ B^{th}\geq 2\, bit/subcarrier $. \par However there is little enhancement than the classical  (OFDM-4QAM) scheme in terms of spectral efficiency, there is a 3dB saving in terms of power efficiency as the average active subcarrier is $\frac{N}{2}$. In other words, the OFDM symbol is sent with half power. 
\subsubsection{Error Performance Analysis}\label{BS-Err}
The derivation of the Symbol error rate of EGIM-4QAM is nearly similar to that derived in  \cite{abu2009SIM} with two important points that will be mentioned later. The detection of symbols constellation shown in Fig.\ref{fig:Bitstuff_Const}  could be divided into two processes:
\begin{itemize}
\item The $ S_o $ symbol detection is exactly like OOK coherent detection.
\item The rest of symbols $\{S_1,S_2,S_3,S_4\} $ will be treated as detection  (4QAM) symbols.
\end{itemize} 
The total probability of symbol error in our scheme is the summation of error in  inactive indices (ook symbol) and the error in active indices symbol (4-qam) as follows:
\begin{eqnarray}\label{eq:BS_ERR}
 P_{sym}&=P(e/ook)P_{ook}+P(e/qam)P_{qam}\\
         &=\frac{1}{2}\dot{P(e/ook)}+\frac{1}{2}\dot{P(e/qam)}
  \end{eqnarray}
  where the number of active subcarriers is a random variable Bernoulli distributed with a mean( $ \frac{N}{2} $),  therefore the probability of active subcarriers within the frame is equal to the inactive subcarriers $P(ook)=P(qam)=0.5 $
  \begin{enumerate}
 \item{The OOK Symbol Error Analysis  }  
 computed by using the closed form of the symbol error rate of the OOK detector is derived as follows:
  \begin{equation}\label{eq:Pook}
  P(e/ook)=\frac{1}{2}\left(-\sqrt{\frac{0.5\bar{\gamma_s}}{1+0.5\bar{\gamma_s}}}\right)
  \end{equation}
 \par The term  $\bar{\gamma_s} $ is the average SNR per subcarrier, such that $\bar{\gamma_s}=|\bar{h_k}|^2\frac{\bar{E_s}}{N_o} $, where $\bar{E_s}  $ and $ \left|\bar{h_k}\right| $ is the average energy per symbol and the channel amplitude respectively. 
 The idea of using half SNR in the above formula is due to the use of half of transmitting power since the inactive indices are sent with no power, the decision region to distinguish between the active and inactive subcarrier is a circle whose center in the origin and radius is equal to the half Euclidean distance between $ S_o $ and any other symbol $ \{S_1,S_2,S_3,S_4\} $.
  \item {The QAM Symbol Error Analysis }
  computed by using the generalized closed form of symbol error rate probability of detecting QAM (M-ary) symbol over fading channel is derived in \cite{alouini1999ERR-OOK}, we can tailor the result for our scheme presented in section \ref{sec: EGIM-1}, also mentioned in \cite{abu2009SIM} as follows:
  \begin{equation}\label{eq:PQAM}
     \begin{split}
  P_{qam}& = -\frac{1}{8}-\frac{1}{2}\sqrt{\frac{\bar{\gamma_s}}{2+\bar{\gamma_s}}}\\&+\frac{1}{2\pi}\sqrt{\frac{\bar{\gamma_s}}{2+\bar{\gamma_s}}}\tan^{-1}\sqrt{\frac{\bar{\gamma_s}}{2+\bar{\gamma_s}}}
   \end{split} 
  \end{equation}
    \end{enumerate}
 The average symbol error probability for EGIM-4QAM scheme introduced in equation (\ref{eq:SER-EGIM1})  derived by substituting with equations (\ref{eq:Pook} and \ref{eq:PQAM}) in (\ref{eq:BS_ERR}). The result is identical to what is derived in \cite{abu2009SIM}. However, many considerations must be taken into account as follows:
 \begin{itemize}
 \item In \cite{abu2009SIM} the author assumes the independence between $ B_{ook} $( index bits) and $ B_{QAM} $ ( symbol mapper bits). However, they are correlated, since the resulting errors in $ B_{ook} $ will spread out across the frame causing the whole frame error, while in our scheme the error is bounded within symbol bounds for the same case.
 \item The excess subcarrier that carries the information about the activation pattern depends on the majority of bits (Zeros or Ones), which is also vulnerable to channel error, which will cause total frame error. 
\end{itemize}   
\subsection{The Second Case study for EGIM-8PSK }\label{sec:EGIM-2}

 In \cite{DM-OFDM-IM} the symbols are drawn from one of two constellations $M_1$ and $M_2$ depending on the activation process. 

\begin{figure}[!t]
\centering
\includegraphics[width=2.8in, height=2.6 in]{ 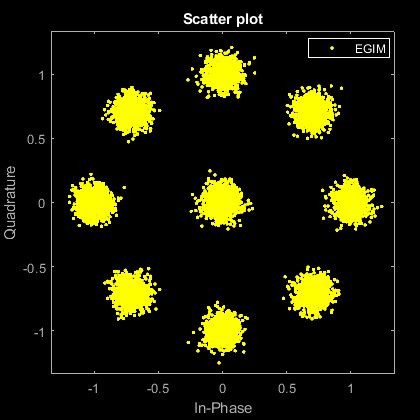}
\caption{Modified constellation points 8PSK  over AWGN, SNR=20dB}
\label{fig:proposed constellation}
\end{figure}

In this section, we will extend the idea of bit-stuffing introduced section \ref{sec: EGIM-1} to be applied across the whole frame of DM-OFDM-IM in \cite{DM-OFDM-IM} using 8psk constellation instead of ($M_A$, $M_B$), with insertion of the origin point (0+0j) to carry the off symbol as shown in Figure \ref{fig:proposed constellation}, the complete symbol Mapping proposed in such case:
\begin{table}[!h]
\caption{proposed Mapper for second case study\label{tab:Mapper}}
\centering
\begin{tabular}{|c||c||c|}
\hline
symbol&code word & constellation\\
\hline
$S_o$&0000& $0+0j$\\

\hline
$S_1$&1000& $1+0j$\\
\hline
$S_2$&1001& $0+1j$\\
\hline
$S_3$&1010& $-1+0j$\\
\hline
$S_4$&1011& $0-1j$\\
\hline
$S_5$&1100& $\dfrac{1}{\sqrt{2}}(1+j)$\\
\hline
$S_6$&1110& $\dfrac{1}{\sqrt{2}}(-1+j)$\\
\hline
$S_7$&1111& $\dfrac{1}{\sqrt{2}}(-1-j)$\\
\hline
$S_8$&1101& $\dfrac{1}{\sqrt{2}}(1-j)$\\
\hline
\end{tabular}
\end{table}
\begin{figure}
    \centering
    \includegraphics[width=2.6in, height=2.6in]{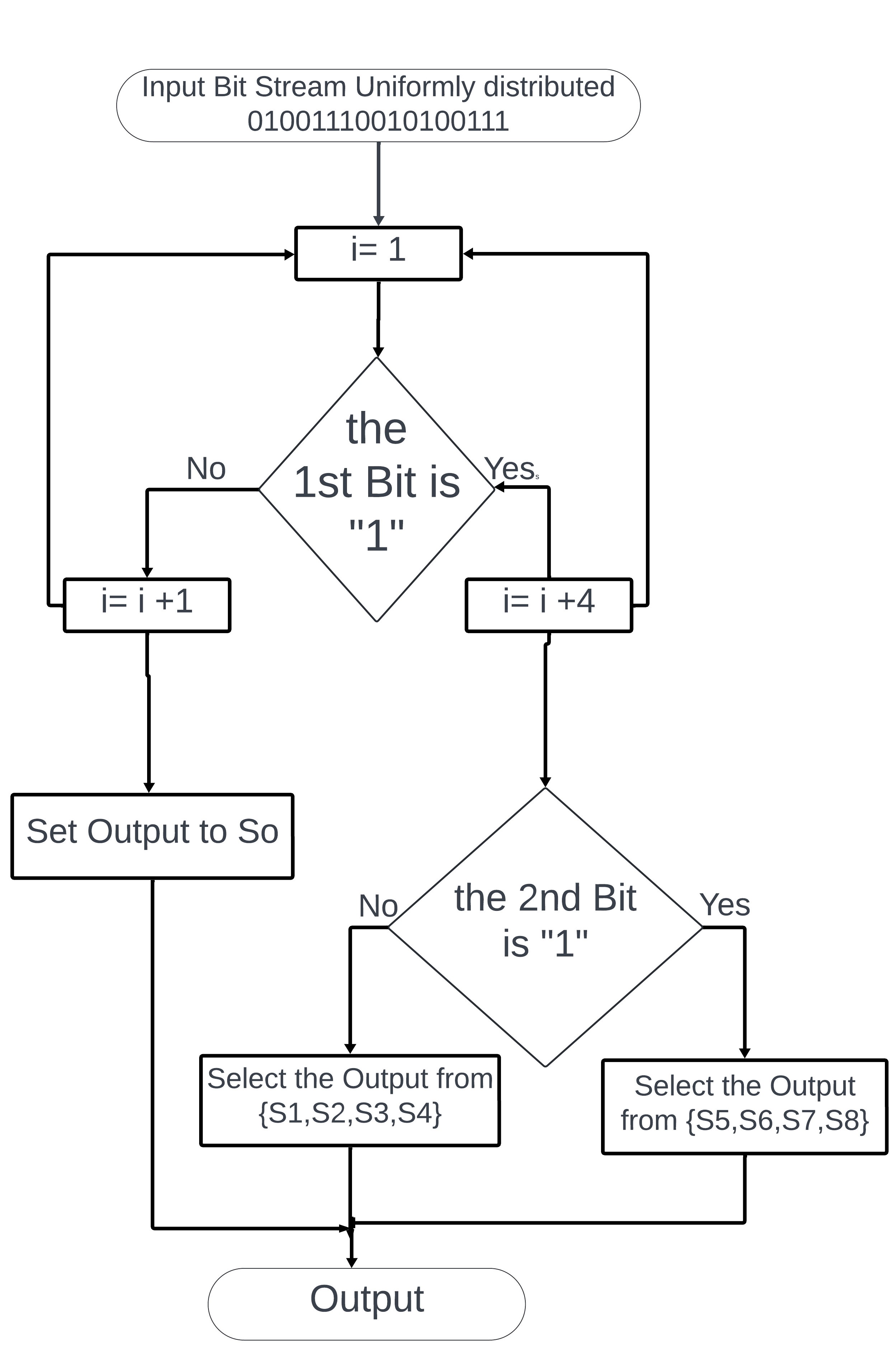}
    \caption{The Generation of EGIM-8PSK}
    \label{fig:EGIM-flow}
\end{figure}
The equivalent transmitter of EGIM  is described in Fig.\ref{fig:Bit_stuffing_TX}  where the input symbols are drawn from the 8PSK constellation.  The  EGIM-8PSK generation process  is depicted in Fig. \ref{fig:EGIM-flow} as follows:
\begin{enumerate}
    \item If the input bit zero is replaced with "0000" the Transmitter jumps the following bits.
    \item If the input bit is "1" the transmitter investigates the following bit:
    \begin{enumerate}
        \item Case "0" the output symbol is selected from the $M_1$ Constellation\\(i.e.  $S\in\{S_1,S_2,S_3,S_4\}$).
        \item Case "1" the output symbol is selected from the $M_2$ Constellation\\(i.e.  $S\in\{S_5,S_6,S_7,S_8\}$).
    \end{enumerate}
\end{enumerate}
Each of the nine codewords is mapped to in-phase and quadrature components as mentioned in table \ref{tab:Mapper}. \\The Major difference between the work introduced in \cite{NEW-DM} is using an 8PSK Constellation with an additional point inserted with zero power representing the off subcarriers that enhance the overall spectral efficiency.

\subsection{Spectral Efficiency Analysis}
In our case, all symbols $S= {S_o,S_1, S_2,....,S_8}$ within the OFDM frame will be represented by  "4bits", the most significant bit in each symbol represents the index pattern, the other 3 bits will depend on the index bit as follows :
\begin{enumerate}
    \item Case "1" the 3 bits will represent constellation points drawn from the 8PSK scheme. 
    \item Case "0" index bit will be followed with "000" to preserve the symbol dimension.
\end{enumerate} 

 Since symbols are drawn from the 8PSK constellation (i.e. M=8), the average spectral efficiency per subcarrier is:
 \begin{equation}
B_{8PSK}\geq 1+\frac{1}{2}\log_2(M).
\end{equation}
 Therefore the  number of bits per subcarrier  will be:
\begin{equation}
B\geq 2.5 bits.
\end{equation}

\subsection{Error Performance Calculation EGIM-8PSK}
In this section, we will derive the necessary formulas for the symbol error rate for EGIM-8PSK. For constellation point $ S_i $ where $ i=0,1,..,M-1. $shown in Fig. \ref{fig:proposed constellation}, each points correspond to a symbols as in Table \ref{tab:Mapper}. The average symbol error rate for the $ k^{th} $ subcarrier, where $ k=1,2,....,N $, can be calculated in the same way as in section \ref{BS-Err} except for using the closed form probability of error of M-PSK instead of the one for M-ary QAM symbols.
The symbol error probability of the off symbol will be given by: 
\begin{equation}\label{eq:EGIM-Pook}
  P(e/ook)=\frac{1}{2}\left(1-\sqrt{\frac{0.5\bar{\gamma_s}}{1+0.5\bar{\gamma_s}}}\right)
  \end{equation}
The closed-form probability of error of M-PSK given in \cite{SER-MPSK} as follows:
 \begin{equation}\label{eq:SER-MPSK}
 \begin{split}
  P_{psk}=\frac{M-1}{M}-\frac{1}{\pi}\sqrt{\frac{\bar{\gamma_s}\sin^2{\frac{\pi}{M}}}{1+\bar{\gamma_s}\sin^2{\frac{\pi}{M}}}}
\\ \Bigl( \frac{\pi}{2}+\arctan\Bigl(\sqrt{\frac{\bar{\gamma_s}\sin^2{\frac{\pi}{M}}}{1+\bar{\gamma_s}\sin^2{\frac{\pi}{M}}}}\Bigl)\cot{\frac{\pi}{M}}\Bigl)
 \end{split}
  \end{equation}

$  $The value of $\gamma$ is equal to $ \frac{1}{N_o} $, where $N_o$ is the variance of noise sample per subcarrier.

\begin{table*}[!t]
\caption{Symbol Error Rate of EGIM }

\begin{minipage}{1.0\textwidth}
\begin{align}
\label{eq:SER-EGIM1}
\hspace{-100pt}
 P_{4QAM}  &=\frac{1}{2}\Bigr[\frac{1}{2}\left(1-\sqrt{\frac{0.5\bar{\gamma_s}}{1+0.5\bar{\gamma_s}}}\right)\Bigr] +\frac{1}{2}\Bigr[ \frac{-1}{8}-(\frac{1}{2})\sqrt{\frac{\bar{\gamma_s}}{2+\bar{\gamma_s}}}
 + \frac{1}{2\pi}\sqrt{\frac{\bar{\gamma_s}}{2+\bar{\gamma_s}}}\tan^{-1}\Bigl({\sqrt{\frac{\bar{\gamma_s}}{2+\bar{\gamma_s}}}}\Bigr)\Bigr]\\
{P}_{8PSK} \label{eq:SER-EGIM2}&= 
\frac{1}{2}\Bigr[\frac{1}{2}\left(1-\sqrt{\frac{0.5\bar{\gamma_s}}{1+0.5\bar{\gamma_s}}}\right)\Bigr]
 +\frac{1}{2}\Bigr[ \frac{7}{8}-(\frac{1}{\pi})\sqrt{\frac{\bar{\gamma_s}\sin^2{(\frac{\pi}{8})}}{1+\bar{\gamma_s}\sin^2{(\frac{\pi}{8})}}}
 \Bigl( \frac{\pi}{2}+\tan^{-1}\Bigl(\sqrt{\frac{\bar{\gamma_s}\sin^2{(\frac{\pi}{8})}}{1+\bar{\gamma_s}\sin^2{(\frac{\pi}{8})}}}\Bigl)\cot{(\frac{\pi}{8}})\Bigl)\Bigr]
\end{align}
\medskip
\hrule
\end{minipage}
\end{table*}

The total average symbol error rate EGIM-8PSK is listed in equations  (\ref{eq:SER-EGIM2}), the reason for computing SER rather than the BER is the huge computational complexity of BER  formulas listed in \cite{BER-IM} due to the large number of indices pattern the receiver detector have to distinguish.  
\\For the Bit error probability  $10^{-3}$, a comparison of the two EGIM case study with different forms of IM in terms of received SNR and spectral efficiency per subcarrier is shown in Table \ref{tab:SE Comp}.\\
\begin{table}[!h]
 \caption{Spectral efficiency comparison}
 \label{tab:SE Comp}
\begin{center}
\fontsize{9pt}{12pt}\selectfont
\begin{tabular}{ | m{2.2cm} |m{1.3cm}| m{2.5cm}|  }
\hline
  scheme & SNR (dB)& $bits/subcarrier$ \\
  \hline
 EGIM-4QAM&24 & $\geq2.5$\\
  \hline
  GIM-2& 25 &2.75\\
  \hline
 DM-OFDM-IM& 25 &2.5 \\
 \hline
 EGIM-8PSK & 23 &$\geq2 $\\ 
  \hline
  GIM-1&23 & 1.222 \\
  \hline
\end{tabular}
\end{center}
\end{table}
The two cases offer a good choice of error performance and spectral efficiency. In contrast, the Enhanced generalized techniques have systematic generation and detection at a complexity almost near the classical OFDM system. However, they are also vulnerable to error propagation when OFF-data symbols are wrongly demapped into the ON-symbol and vice versa. \\The solution to this problem will be presented in the next section \ref{sec:auto_enc}, where the error propagation across OFDM-IM frames is avoided by exploiting the error correction capabilities of the proposed OFDM-IM  convolutional autoencoder. 
\section{Implementation of OFDM-IM using Autoencoder}\label{sec:auto_enc}
This section will demonstrate the generation of error propagation-free OFDM-IM using a customized convolutional encoder, this method will systematically convert the input binary stream into on and off symbols. Additionally, the autoencoder error capabilities prohibit the event of error propagation mentioned in section \ref{sec:EGIM-2}.
\subsection{Auto Encoder Structure}
\begin{figure}[!ht]
    \centering
    \includegraphics[width=3.0in, height=1.7in]{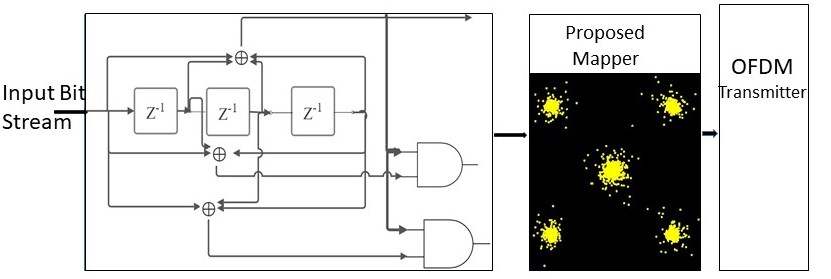}
    \caption{The structure of OFDM-IM autoencoder}
    \label{fig:Conv-enc}
\end{figure}
\begin{figure}[!t]
\centering
\includegraphics[width=3.2in,height= 2.6in]{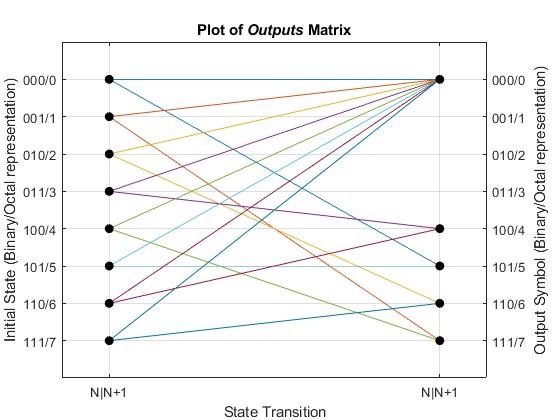}
\caption{The output trellis of the OFDM-IM autoencoder }  
\label{fig:trellis}
\end{figure}

The proposed autoencoder is depicted in Fig. \ref{fig:Conv-enc} as a simple convolutional encoder with rate $\frac{1}{3}$ and constraint length of order 4, where its  generator polynomials  will be as follows:
\begin{align*}
    g_1(x)&=x^3+x^2+x+1\\
    g_2(x)&=(x^3+x^2+x).g_1(x)\\
    g_3(x)&=(x^3+x+1).g_1(x)\\
\end{align*}
The $1^{st}$ output is mixed with the $2^{nd}$ and $3^{rd}$ with the mean of an And gate which will force the output to all zero symbol (i.e. off symbol) in case the zero is out of  $1^{st}$ generator, On other hands when the  $1^{st}$ output is "1" (i.e. active) allows the encoder to send the constellation symbols.
There is a need to define a new term called the effective symbol rate to determine the compared candidate-coded system benchmark as follows:
\begin{equation}
    Sym_{eff}= code rate \times \log_2M.
\end{equation} Where the $Sym_{eff}$ of our proposed system (Encoder, Mapper) is "1".
The resultant output trellis structure is shown in Fig.\ref{fig:trellis}, the encoder systematically maps the input bit stream into five codewords listed in Table\ref{tab:Bitstuff Mapper}. Where, $S_o$ represents the inactive symbol and the symbols $S_{on}={S_1,S_2,S_3,S_4}$ are the active symbols. Hence, the resultant  probabilities of the output symbols will be:
\begin{equation}
 P_{S_o}=P_{S_{on}}=\frac{1}{2} 
 \end{equation} 

 For the input bit stream
 $[m=m_1,m_2,....,m_i]$, the coded output will be:
 \begin{equation}
  S_i=G(m_i) \; where \; S_i\in [S_0,...,S_4]
 \end{equation}
 In the receiver, we employ the Viterbi decoding algorithm with either hard or soft decisions as follows:
 \subsection{Hard Decision Decoding} 
 The output bit stream results from the decoder  using hard decision Viterbi algorithm follows:
\begin{equation}
d_H=\sum_{i=1}^{tb}|c_i-b_i|
\end{equation}
Where:
$d_H$ is the Hamming distance,
$c_i$ is the received bit,
$b_i$ is the expected bit and 
 tb is the traceback length (i.e.tb =3x constraint length ).\\
 \par  The Viterbi algorithm checks the Hamming distance for all possible paths and selects the route with the minimum distance. Hard decision decoding relies on binary decisions (0 or 1) and compares these to the expected bits to identify and correct errors.
  \subsection{Soft Decision Decoding}
 The decoder input will be samples computed  using approximate LLR algorithm in \cite{Soft-Dec} instead of binary input  as follows:
 \begin{multline}
    \mathcal{L}(b)=\frac{-1}{\sigma^2}\Bigl(\min_{s \in S_o}((x-s_x)^2+(y-s_y)^2)\\-\min_{s \in S_1}((x-s_x)^2+(y-s_y)^2)\Bigr)  
  \end{multline}
 Where $x+jy$ is the received signal point, $\sigma^2$ is the inband  noise variance, b is one of the k bits of (M-ary) symbol and
     $s_x,s_y$  is the in-phase and quadrature component of the true constellation points.
    \\In soft decision the Euclidean distance is used as a metric of the decision of the survivor branch on the trellis instead of Hamming distance used in hard decision.
\section{Simulation Results}
\label{sec:sim_res}
 \begin{table}[!h]
\caption{Simulation Parameters\label{tab:Sim_Par}}
\centering
\fontsize{10pt}{12pt}\selectfont
\begin{tabular}{ || m{2.8cm} | m{3.8cm}|| }
\hline
Parameters &  value\\
\hline
Framework & Matlab 2024Ra \\
\hline
Mapping &4QAM, 8PSK \\
\hline
FFT&64 \\
\hline
cyclic prefix&16\\
\hline
channel & Rayleigh, AWGN(coded)\\
\hline
number of taps&10 \\
\hline
equalizer& MMSE\\
\hline
detector& ML\\
\hline
Traceback Length& 12\\
\hline
Constraint Length& 4\\
\hline
\end{tabular}
\end{table}

In this section, we will deduce Matlab simulation results for what had driven in sections \ref{sec: EGIM-1} and \ref{sec:EGIM-2}, the error performance for EGIM-4QAM and  EGIM-8PSK will be demonstrated in terms of Symbol Error Rate(SER), compared with their theoretical SER  upper bound, also the bit error rate of the OFDM-IM autoencoder. 
  The performance of hard and soft decision Viterbi algorithm of the autoencoder will be compared with a standard convolutional encoder of rate $\frac{1}{2}$ produces the same number of symbols.    

\subsection{Symbol Error Rate of  EGIM-4QAM}\label{Err_BS}
\begin{figure}[!t]
\centering
\includegraphics[width=3.0in]{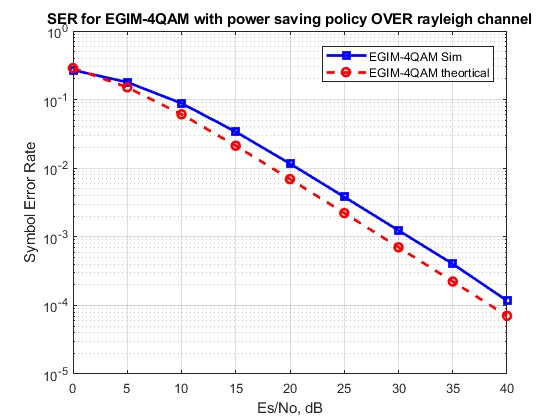}
\caption{SER of EGIM-4QAM Verus theoretical bound at SE$\,\geq$ 2bit  }
\label{fig:Bit_stuf}
\end{figure}
The simulation in Fig.\ref{fig:Bit_stuf} shows the difference between the theoretical upper bound SER driven in equation (\ref{eq:SER-EGIM1}) and simulation of EGIM-4QAM, where the error event introduced will localize within the symbol bounds while involving all possible combination patterns results in higher spectral efficiency as stated in section \ref{sec:SE_BS}.  
\subsection{ Error Performance of  EGIM-8PSK}\label{Err_EGIM} 
\begin{figure}[!ht]
\centering
\includegraphics[width=3.0in]{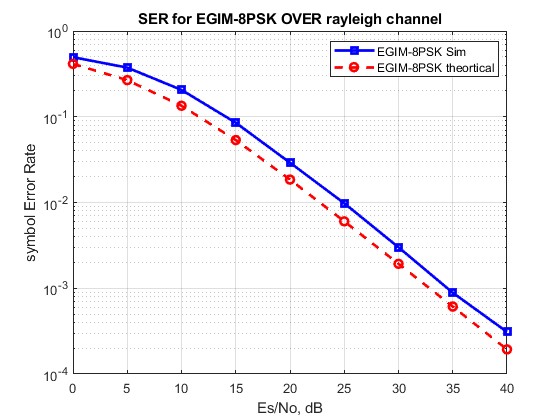}
\caption{SER of EGIM-8PSK Verus theoretical bound at SE$\,\geq$ 2.5bit}
\label{fig:EGIM-8psk}
\end{figure}

\begin{figure}[!t]
\centering
\includegraphics[width=3.0in, height=2.5in]{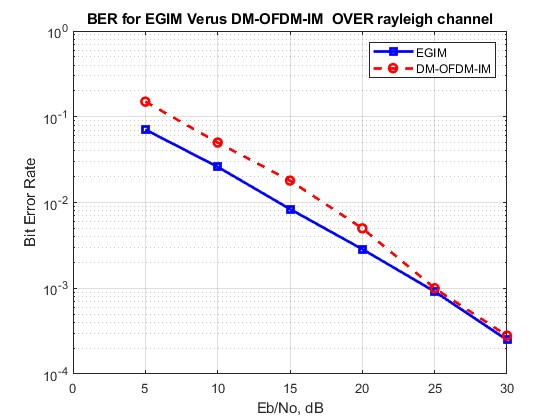}
\caption{BER of EGIM-8PSK Verus DM-OFDM-IM at SE$\,\geq$ 2.5bit}
\label{fig:EGIM-DM}
\end{figure}

The validation of EGIM-8PSK SER theoretical upper bound stated in equation (\ref{eq:SER-EGIM2}) with simulation is shown in  Fig. \ref{fig:EGIM-8psk}. Where in Fig.\ref{fig:EGIM-DM}, the BER rate of EGIM-8PSK is compared with DM-OFDM-IM introduced in \cite{DM-OFDM-IM} for the same spectral efficiency (2.5 bits/subcarrier). The comparison revealed the superiority of EGIM-8PSK especially over the low SNR range. 
The validation of   Fig. \ref{fig:EGIM-8psk}. shows the SER validation between  EGIM-8PSK and its theoretical upper bound stated in equation (\ref{eq:SER-EGIM2}). Where in Fig.\ref{fig:EGIM-DM}. the BER rate of EGIM-8PSK is compared with DM-OFDM-IM introduced in \cite{DM-OFDM-IM} for the same spectral efficiency (2.5 bits/subcarrier). The comparison revealed the superiority of EGIM-8PSK especially over the low SNR range. 

%\begin{figure}[!t]
%\centering
%\includegraphics[width=3.0in]{images/saving.jpg}
%\caption{Soft and Hard decoding with power saving policy over AWGN}
%\label{fig:power saving}
%\end{figure} 
\subsection{Performance of proposed Encoder}
\begin{figure}[!ht]
\centering
\includegraphics[width=3.0in]{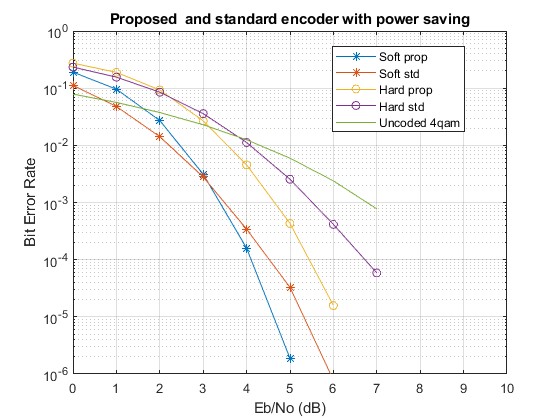}
\caption{The Error Performance of OFDM-IM autoencoder with power reinvesting policy over AWGN}
\label{fig:power reinvesting}
\end{figure} 

\begin{figure}[!ht]
\centering
\includegraphics[width=3.0in]{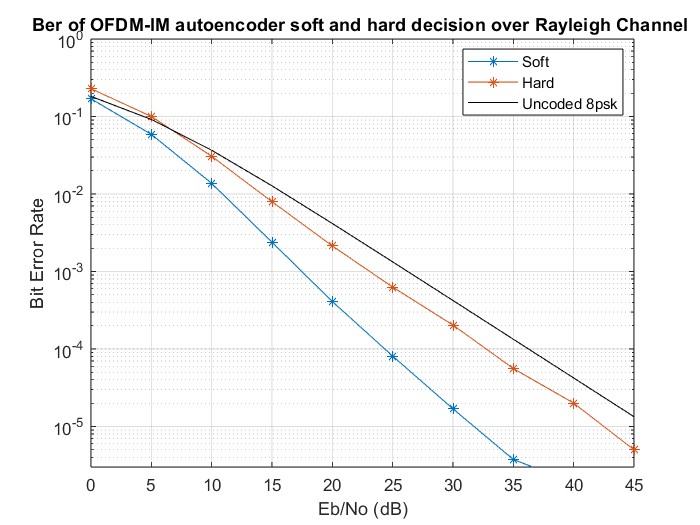}
\caption{Ber of OFDM-IM autoencoder soft and hard decision over Rayleigh Channel}
\label{fig:Rayleigh}
\end{figure} 
When compared to a system with the equivalent  $Sym_{eff}  (R=\frac{1}{2}, 4QAM)$ the proposed convolutional encoder will offer (3dB) enhancement in terms of power efficiency since approximately half of OFDM-IM frame subcarriers are inactive (i.e.: sent with zero power) according to the allowable output symbols of the encoder shown in Fig.\ref{fig:trellis}.\\
The 3dB saved power can used in two different ways as follows:
\begin{enumerate}
    \item It can be reinvested into the active subcarrier in the OFDM-IM frame, resulting in enhancement of the error performance (0.8dB) in case of soft decision Viterbi decoding while the enhancement will (1.2dB) in the hard decision when compared to a standard encoder of rate $\frac{1}{2}$ as shown in Fig.\ref{fig:power reinvesting}.
    \item Achieving the same error probability with higher order modulation improves spectral efficiency.
\end{enumerate}
The performance of the OFDM-IM autoencoder is tested over the Rayleigh channel using hard and soft decision decoding demonstrated in Fig.\ref{fig:Rayleigh}. these results make sure of the autoencoder's robustness against the channel conditions.
The simulations show that the proposed encoder eliminates the chance of the error propagation phenomena that occur when using index modulation with variable active indices over high-order modulation.

\section{Conclusion and Future Work}
\label{sec: conclusion}
A novel proposed technique for systematically generating and detecting OFDM-IM has been presented. The effect of this technique is studied through two case studies (4QAM-8PSK). Making full use of the whole span over the OFDM-IM frame without splitting reducing transmitter complexity, and involving all possible combination patterns results in enhancement in spectral efficiency per subcarrier without introducing extra computational complexity into the receiver ML detector.  The symbol error rates upper bound for (4QAM-8PSK) were driven, and the simulation of both techniques converage their theoretical bounds.
An error propagation OFDM-IM autoencoder was presented to generate pattern combinations in addition to the transmitted bit stream through a special design of a convolutional encoder. The nonlinearity added to the convolutional encoder to generate active and inactive symbols has a 3dB power efficiency gain at the expense of the encoder-free distance which affects its overall error performance.

In the future, one of the main challenges is to enhance the performance of the OFDM-IM autoencoder, achieving almost the standard coding error performance while maintaining the 3dB power efficiency gain or using different channel coding methods to achieve the same concept.   

\section*{Acknowledgments}
The authors would like to thank the anonymous reviewers
for valuable comments that have led to improvements in this
paper.


\bibliographystyle{IEEEtran} 
\bibliography{References.bib}
\end{document}